\documentclass[12pt,preprint]{aastex}

\newcommand{\kms}{km s$^{-1}$}

\shorttitle{Trigonometric Parallaxes: G9.62+0.20}
\shortauthors{Sanna et al.}

\begin{document}

\title{TRIGONOMETRIC PARALLAXES OF MASSIVE STAR-FORMING REGIONS. VII. G9.62+0.20 AND THE EXPANDING 3~KPC-ARM}

\author{A. Sanna\altaffilmark{1,2,3}, M. J. Reid\altaffilmark{3}, L. Moscadelli\altaffilmark{4},
T. M. Dame\altaffilmark{3}, K. M. Menten\altaffilmark{5}, A. Brunthaler\altaffilmark{5},
X. W. Zheng\altaffilmark{6} and Y. Xu\altaffilmark{7}}

\email{asanna@ca.astro.it}
\altaffiltext{1}{Dipartimento di Fisica, Universit\'a degli Studi di Cagliari, S.P. Monserrato-Sestu km 0.7, I-09042 Cagliari, Italy}
\altaffiltext{2}{INAF, Osservatorio Astronomico di Cagliari, Loc. Poggio dei Pini, Str. 54, 09012 Capoterra (CA), Italy}
\altaffiltext{3}{Harvard-Smithsonian Center for Astrophysics, 60 Garden Street, Cambridge, MA 02138, USA}
\altaffiltext{4}{INAF, Osservatorio Astrofisico di Arcetri, Largo E. Fermi 5, 50125 Firenze, Italy}
\altaffiltext{5}{Max-Planck-Institut f\"{u}r Radioastronomie, Auf dem H\"{u}gel 69, 53121 Bonn, Germany}
\altaffiltext{6}{Department of Astronomy, Nanjing University, Nanjing 210093, China}
\altaffiltext{7}{Purple Mountain Observatory, Chinese Academy of Sciences, Nanjing 210093, China}

\begin{abstract}
We report a trigonometric parallax of 12~GHz methanol masers associated with the massive star forming region \objectname{G9.62+0.20},
corresponding to a distance of $5.2^{+0.6}_{-0.6}$~kpc.  With a local standard of rest velocity of about 2~\kms, the region's kinematic
distances of 0.5 and 16~kpc differ greatly from the distance derived here. Our measurement of the peculiar motion
of the star forming region shows a very large deviation from a circular Galactic orbit: 41~\kms\
radially outward from the Galactic center and 60~\kms\ counter to Galactic rotation. The combination of its
radial velocity and distance places \objectname{G9.62+0.20} in the inner region of the Galaxy close to
the Expanding Near 3~kpc-Arm, where the bulge/bar potential has strong gravitational influence.
We also map the distribution of 12~GHz methanol masers, locate them with respect to a hypercompact H~II
region, and compare our data with the periodic flare phenomenon reported previously for this source.

\end{abstract}

\keywords{astrometry --- Galaxy: fundamental parameters --- Galaxy: kinematics and dynamics
--- masers --- stars: formation --- techniques: high angular resolution}

\section{Introduction}

\objectname{G9.62+0.20} is a massive star-forming complex harboring a number of H~II regions in different
evolutionary phases (e.g. \citealt{Garay1993,Hofner1996,Testi2000,Hofner2001,DeBuizer2003}).
Several masing species are associated with component \objectname{G9.62+0.20}~E, which has been classified as
a candidate hypercompact (HC) H~II region \citep{Kurtz2002}.  Both 6.7 and 12.2 GHz methanol masers have been
detected toward this object (\citealt{Phillips1998,Caswell1995}), as well as 22.2~GHz water \citep{HandC1996}
and 1.665 and 1.667~GHz hydroxyl \citep{Caswell1998} masers. Furthermore \objectname{G9.62+0.20} has been
claimed to be the most luminous 6.7~GHz methanol maser known in the Galaxy and both its 6.7 and 12.2~GHz
methanol emissions show periodic flaring (\citealt{Goedhart2003,Goedhart2005}).

\objectname{G9.62+0.20} is in the first Galactic quadrant, projected 9.6 degrees from the Galactic center.
While its Local Standard of Rest (LSR) velocity is near zero, 21-cm wavelength H~I spectra toward
its H~II region(s) show clouds of absorbing gas with LSR velocities out to 55~\kms \citep{Hofner1994}. This
effectively rules out the near kinematic distance of 0.5~kpc \footnote{We adopt a ``revised'' kinematic distance
using the prescription of \citet{Reid09b}, assuming \objectname{G9.62+0.20} has an LSR velocity of +2.0~\kms.}.
The apparent cutoff in H~I absorption above 55~\kms\ suggests that the source is not at the far distance either
(16~kpc), but rather lies deep in the inner Galaxy on the near side of the tangent point.
However, this argument may partly be affected by the decrease of cold and dense H~I inside the Galactic
gas hole (Galactocentric radius $\rm{R}<3$~kpc), since this gas component is the one responsible for well-defined
H~I absorption lines.

In order to determine the true distance and reliably locate \objectname{G9.62+0.20} in the Galaxy,
we conducted multi-epoch, phase referencing, VLBA observations of its 12~GHz methanol masers.
In this paper, we report a measurement of trigonometric parallax based on our spectral line interferometric
observations. In addition to the parallax, we accurately determine the proper motion of the masers with
respect to background quasars.  Combining the coordinates, distance, LSR velocity and proper motions
yields the full 3-dimensional location and kinematics of the massive star-forming region.
This complete information clearly associates \objectname{G9.62+0.20} with the complex inner region of
the Galaxy where the central bar has strong gravitational influence.

\section{Observations and Data Analysis}

Using the NRAO\footnote{The National Radio Astronomy Observatory is a facility of the National
Science Foundation operated under cooperative agreement by Associated Universities, Inc.}
Very Long Baseline Array (VLBA), we observed the high-mass star forming region (HMSFR)
\objectname{G9.62+0.20} in the $2_0-3_{-1}$~E CH$_3$OH transition
(rest frequency 12178.597~GHz).
Background information about our VLBA program to measure parallaxes for HMSFRs
is given in \citet{Reid09}, hereafter called Paper I.
We observed under program BR100A at 5 epochs:
2005 March 25 and October 8, 2006 March 16 and October 4, 2007 March 16.
These dates were chosen to sample the peaks of the parallax signature in Right Ascension, since
the amplitude of the Declination parallax signature is small and those position measurements are
more susceptible to systematic errors at the low Declination of this source.

\begin{figure*}
\centering
\includegraphics[angle= -90, scale= 0.6]{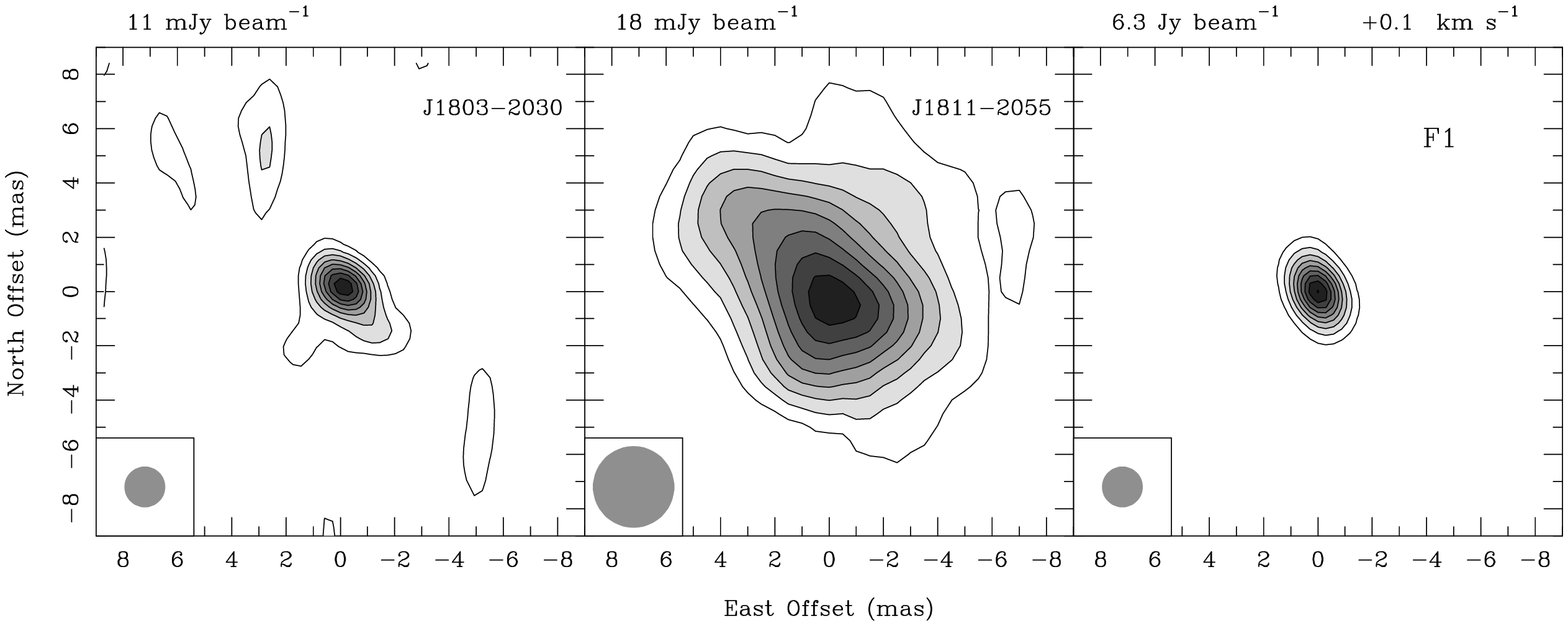}
\caption{Images of the background continuum sources near G9.62+0.20 and the phase reference spot F1. Source names are in the upper right corner and restoring beams are in the lower left corner of each panel. Peak intensities are reported on the top of each panel as well the V$_{LSR}$ of the spot F1. Contour levels are at multiples of 10\% of each peak. All images are from the third epoch observations on 2006 March 16. \label{fig1}}
\end{figure*}

Paper I describes the general observational strategy and data calibration procedures.
The maser observations were centered at the systemic velocity
(V$_{LSR}$) of +2.0~\kms, also traced with the CH$_3$CN~$(6-5)$ line emission by \citet{Hofner1996}.
Spectral resolution was 0.38~\kms.
Two fringe finders, J1743--0350 and J1800+3848 from the ICRF catalog, were observed for single-band
delay and instrumental phase-offset calibration. For background position references, we
observed 4 calibrators, J1803--2030, J1804--2147, J1808--2124 and J1811--2055, from the extragalactic
radio survey by \citet{Xu2006}.  Only the first and last calibrators were detected with our VLBA
measurements (see Table~\ref{tab1} and Figure~\ref{fig1}).
We selected the LSR velocity channel at +0.1~\kms\ as the phase reference, corresponding
to the peak intensity of the bright and compact maser spot F1, following the labeling of
\citet{Minier2002} (see Figure~\ref{fig1} and~\ref{fig3}). Table~\ref{tab1} summarizes source properties.

The absolute positions of the two background sources were not known with sufficient accuracy to
be used to determine the absolute position of the maser emission.  However, the {\it a priori}
position of the maser reference spot (F1) was correct to about 20~mas, as indicated by the
low fringe rates of the interferometer phases.
After calibration, we imaged the phase reference maser channel and continuum sources using the
AIPS task IMAGR with a circular restoring beam of $1.5 \times 1.5~\rm{mas}^2$ at each epoch,
equal to the east-west naturally weighted dirty beam size, which was typically
$ 3.8 \times 1.5~\rm{mas}^2 $ at a P.A. of $ 1\degr $ east of north,
since the parallax information comes essentially from the east-west position shifts.
The intensity peak of the phase reference spot varied by less than $\pm20$\% across our
observations spanning 2~yr (see Figure~\ref{fig4}).

Figure~\ref{fig1} shows the images of the background sources, J1803--2030 and J1811--2055, and the phase
reference maser spot from the middle epoch.
For J1803--2030 the small deviations from a pointlike image are most likely caused by unmodeled
short-term atmospheric fluctuations and follow the dirty beam shape (cf. \citealt{Brunt2009}).
The calibrator J1811--2055 turned out to have an elongated-structure, with a possible extended
halo.  When imaging this source, we used a larger circular restoring beam of
$ 3.0 \times 3.0~\rm{mas}^2 $.  This beam size was set to maximize the brightness sensitivity
in the image.  For parallax purposes, we fitted only the peak of the emission.
At the first epoch (2005 March 25) only 4 (BR, FD, NL, OV) VLBA antennas were usable and the
dynamic range of the images was about 7 times worse than for the subsequent 4 epochs.
This precluded the use of the extended source J1811--2055 at this epoch.
We measured the positions of the maser and the background sources by fitting elliptical Gaussian
brightness distributions to the detected emission (AIPS task JMFIT).

\section{Results}

We measured the parallax and the proper motion of 12~GHz methanol masers by modeling the
position differences between the reference maser spot and each calibrator versus time.
The model is the sum of the parallax sinusoid, including the effects of the
ellipticity of the Earth's orbit, and a linear motion in each coordinate.
Since systematic errors usually dominate over random noise, we adopted an ``a posteriori''
estimate of the errors from the fit itself. We assigned independent ``error floors'' to the
east and north position offsets and added them in quadrature to the formal position fitting
uncertainties. This procedure was iterated, adjusting error floors to achieve a reduced
$\chi^2$ per degree of freedom near unity in each coordinate (see Paper~I for a detailed discussion).
We down-weighted the data from the first epoch (2005 March 25) by a factor 7, equal to the
dynamic range degradation of the background source images. We can also put an upper limit on the shift
over time of the centroid position of the quasars due to a possible jet component evolution.
The small residuals for the parallax fits relative to the two quasars, and the similar parallax results,
suggest that such effects are less than about 50~$\mu$as.

\begin{figure*}
\includegraphics[angle=-90,scale=.6]{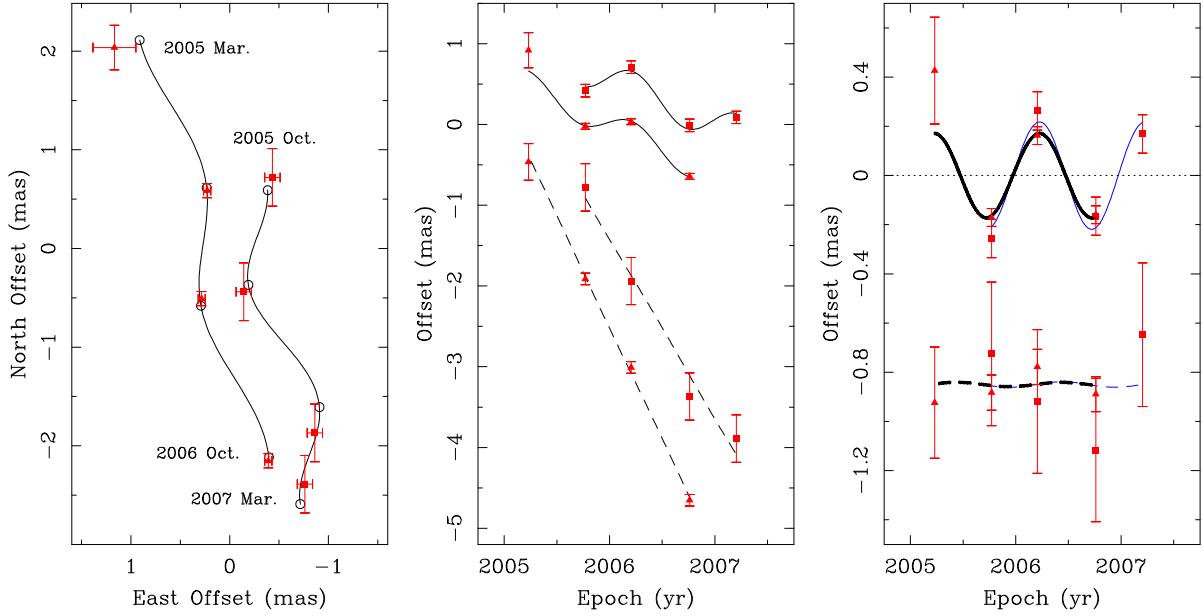}
\caption{ Results of the parallax fit for G9.62+0.20. The different symbols refer to measurements relative to background sources J1803-2030 (triangles) and J1811-2055 (squares). \textit{Left Panel:} Sky projected motion of the maser with respect to J1803-2030 (left) and J1811-2055 (right). The empty circles and the lines show the best-fit position offsets and the trajectory, respectively. \textit{Middle Panel:} The position offsets of the maser along the East and North directions versus time. The best-fit model in East and North direction are shown as continuous and dashed lines, respectively. \textit{Right Panel:} Same as the middle panel but with fitted proper motions subtracted (parallax curve). Parallax curves relative to J1803-2030 (black-thick) and J1811-2055 (blue-thin) are superimposed. The North offset data have been shifted for clarity.
See the electronic edition of the Journal for a color version of this figure.\label{fig2}}
\end{figure*}

Table~\ref{tab2} and Figure~\ref{fig2} show the results of the parallax and proper motion fit
for each background source.  While the post-fit residuals for the fit using J1803--2030 were
about half that of the fit using J1811--2055, the J1803--2030 fit is dominated by data from
only three epochs.  As such, there are effectively near-zero degrees of freedom and the formal
errors are not reliable.   Since we could not reliably determine the relative weights for
the two parallax data sets, we calculated an un-weighted average, instead of the
combined fit procedure applied in previous papers (e.g. Paper~I). The measured parallax of
\objectname{G9.62+0.20} obtained in this manner is $0.194\pm0.023$~mas, corresponding to a distance of
$5.2^{+0.6}_{-0.6}$~kpc.

Our new distance indicates that the luminosity of the brightest 6.7~GHz methanol maser
observed in the Galaxy has been substantially underestimated in the past. 
The near kinematic distance (2.0~kpc) adopted by \citet{Phillips1998} is only 38\%
of our measured distance implying a 7 times brighter source.
This kinematic distance was derived by \citet{F&C1989} based on OH masers
velocities, assuming the \citet{Schimdt1965} Galactic rotation model and a
Sun--Galactic center distance (R$_0$) of 10~kpc.
Adopting the IAU value of the distance to the Galactic center, R$_0 = 8.5$~kpc, our measured distance translates
to a Galactocentric radius of 3.5~kpc. The measured mean proper motion is $-0.580 \pm 0.054 $~mas~yr$^{-1}$
toward the east and $-2.49 \pm 0.27 $~mas~yr$^{-1}$ toward the north;
at our measured distance these values correspond to $-14$~\kms and $-61$~\kms\ eastward and northward, respectively.
Completing the kinematic information, we assume a LSR velocity of $2\pm5$~\kms\ for the maser complex.

We compare our measurements
with a simple rotation model of the Galaxy, converting these values from the equatorial
heliocentric reference frame to a rotating Galactic reference frame. This approach subtracts
a circular velocity component at the position of the source and yields a peculiar motion vector
with respect to circular rotation \citep{Reid09b}. Assuming negligible internal motions of
CH$_3$OH masers (e.g. \citealt{Mosca2002,Mosca2009}) and adopting the IAU value for the
distance to the Galactic center (R$_0 = 8.5$~kpc), the rotation velocity at the solar circle ($\Theta_0 = 220$~\kms),
and the \emph{Hipparcos} measurements of the solar motion \citep{Dehnen1998}, the peculiar velocity
components of \objectname{G9.62+0.20} are $(U_s, V_s, W_s)=( -31 \pm 15, -58 \pm 15, -10 \pm 8)$~\kms, where U$_s$, V$_s$
and W$_s$ are directed toward the Galactic center, in the direction of Galactic rotation and
toward the North Galactic Pole, respectively. Using instead the best-fit values of R$_0 = 8.4$~kpc
and $\Theta_0 = 254$~\kms, obtained from 16 parallax and proper motion measurements \citep{Reid09b},
the peculiar velocity components are $(U_s, V_s, W_s)=( -41 \pm 18, -60 \pm 17, -10 \pm 8)$~\kms.
Quoted uncertainties are based on proper motion and $V_{\rm LSR}$ measurement errors and also a 7~\kms\ uncertainty in each
peculiar velocity component, owing to expected Virial motions of a star within a high mass star forming region.

\section{Galactic Location and Peculiar Motion}

In the literature there are two estimates of distance to the source \objectname{G9.62+0.20}
consistent with our direct measurement.
\citet{Scoville1987} derived a distance of 4.7~kpc for the H II complex \objectname{G9.62+0.20},
based on the association to a CO cloud with a LSR velocity of +4~\kms\ and
following the expanding ring model of the 3~kpc-Arm given by \citet{Bania1980}.
\citet{Hofner1994} showed that there is HI absorption over the LSR velocity range of -5 to 55~\kms\
and suggested a kinematic distance of 5.7~kpc.
Our trigonometric parallax places \objectname{G9.62+0.20} at a distance close to 5.2 kpc.
The combination of distance and Galactic coordinates locates \objectname{G9.62+0.20} in the inner region of
the Galaxy within the so-called ``molecular ring'', which appears as a complex structure in CO
longitude-velocity maps (e.g. \citealt{Dame2001,Rodriguez2008}). Based
on the observations of distinct tangents points, traced by the early CO surveys in the
southern Milky Way  (e.g. \citealt{Robinson1984,Bronfman1989}), the inner edge of
the molecular ring is thought to be composed of the Norma and Expanding 3~kpc arms
(e.g. Figure~5 in \citealt{Bronfman2000}; Figure~3 in \citealt{Dame2001}).
Due to observational difficulties detecting and following the Norma Arm in CO emission
at positive longitudes, the question to which arm, if any, does  the Norma Arm connect in the
first quadrant is still controversial (e.g. \citealt{Russeil2003,Vallee2008}).
Therefore in the following, we focus our comparison of the spatial and velocity properties
of \objectname{G9.62+0.20} with the well-studied Expanding 3~kpc-Arm.

Since the LSR velocity for a nearly circular Galactic orbit at the longitude and distance
of \objectname{G9.62+0.20} would be $+53$~\kms, about 50~\kms\ greater than observed,
\objectname{G9.62+0.20} is clearly highly kinematically anomalous.
We can gain a more complete picture of the region by comparing the Galactic location ($\ell$,\textit{b}) and the LSR
velocity (+2~\kms) of \objectname{G9.62+0.20} with respect to the distribution of CO toward this region.
Using the longitude-velocity and latitude-velocity CO~$(1-0)$ line maps of \citet{Bitran1997}, we find
the HMSFR \objectname{G9.62+0.20} associated with the edge of a CO clump (cutoff at 0.5~K).
\citet{Dame2008} locate the Expanding (Near) 3~kpc-Arm (hereafter simply the 3~kpc-Arm) below the
plane of the Milky Way at Galactic longitude $9.6\degr$, whereas \objectname{G9.62+0.20} is slightly above
the plane.  The LSR velocity of the 3~kpc-Arm in this direction is expected to be $-13$~\kms, blueshifted by
about $53$~\kms\ with respect to a circular orbit at $+40$~\kms.
We find \objectname{G9.62+0.20} to have a $V_{\rm LSR}\approx+2$~\kms,
which, while close to the CO velocity at this longitude, is shifted by about 15~\kms\ with respect
to CO in the 3~kpc-Arm. Thus, \objectname{G9.62+0.20} may not be part of the 3~kpc-Arm as previously proposed and,
perhaps, could belong to a nearby structure such as the Norma Arm. Either way, this source, and presumably
its associated arm, has a large peculiar motion with a $41$~\kms\ component radially outward and a $60$~\kms\
component counter to Galactic rotation, rendering kinematic distances useless.

The spatial and velocity properties of the HMSFR \objectname{G9.62+0.20}
and the 3~kpc-Arm, although not strictly associated, suggest they could participate in a similar dynamical
anomaly. The large peculiar motion of the 3~kpc-Arm might be induced by the gravitational
influence of a central bar(s) potential, as explored by many authors.
\citet{Habing2006} studied the distribution of maser stars in the inner Milky Way with respect to the
ISM distribution and compared their spatial properties with a set of orbits derived by combining
an axisymmetric potential plus a weak rotating bar.
They found an overall agreement between maser stars and gas in both space and velocity,
moving inward along the Galactic plane with almost circular orbits
that become more and more elongated inside the molecular ring with radial motion increasingly important.
Following \citet{Sevenster1999}, they suggested the 3~kpc-Arm to be the locus of ballistic orbits originating
at the corotation radius of the bar.
On the contrary, \citet{Rodriguez2008} modeling the dynamics of the Milky Way gas flow in the presence of
a potential from two nested bars, suggested by the Two Micron All Sky Survey (2MASS) data,
reproduced the longitude-velocity features of the 3~kpc-Arm as a spiral density wave maximum
surrounding the primary bar.
Further parallax and 3-D velocity measurements of HMSFRs sampling the inner region of the Galaxy will help to constraint
the dynamics of this complex region.

As a final remark, we note that two supernova remnants (SNRs), G9.8+0.6 and G9.7-0.0, are projected
to within a half degree of \objectname{G9.62+0.20}. G9.8+0.6 has distance estimates of
8.8~kpc \citep{Stupar2007} and 10.4~kpc \citep{Case1998}, both
based on the $\Sigma$-D relation, which may not be reliable (e.g. \citealt{Green2004}),
whereas the G9.7-0.0 has a near kinematic distance estimate (4.7~kpc) based on an associated OH maser
velocity (+43~\kms) detected toward the SNR \citep{Hewitt2009}.
However, the association of \objectname{G9.62+0.20} and G9.7-0.0 is questionable given
their about 40~\kms\ different radial velocity.

Y.X. was supported by Chinese NSF through grants NSF 10673024,
NSF 10733030, NSF 10703010, and NSF 10621303.

\acknowledgments
We are very grateful to Testi and collaborators for providing the VLA continuum map
of G9.62+0.20.

{\it Facilities:} \facility{VLBA}.

\appendix

\section{Appendix material: 12~GHz Methanol Masers}

\begin{figure*}
\includegraphics[angle=0,scale= .8]{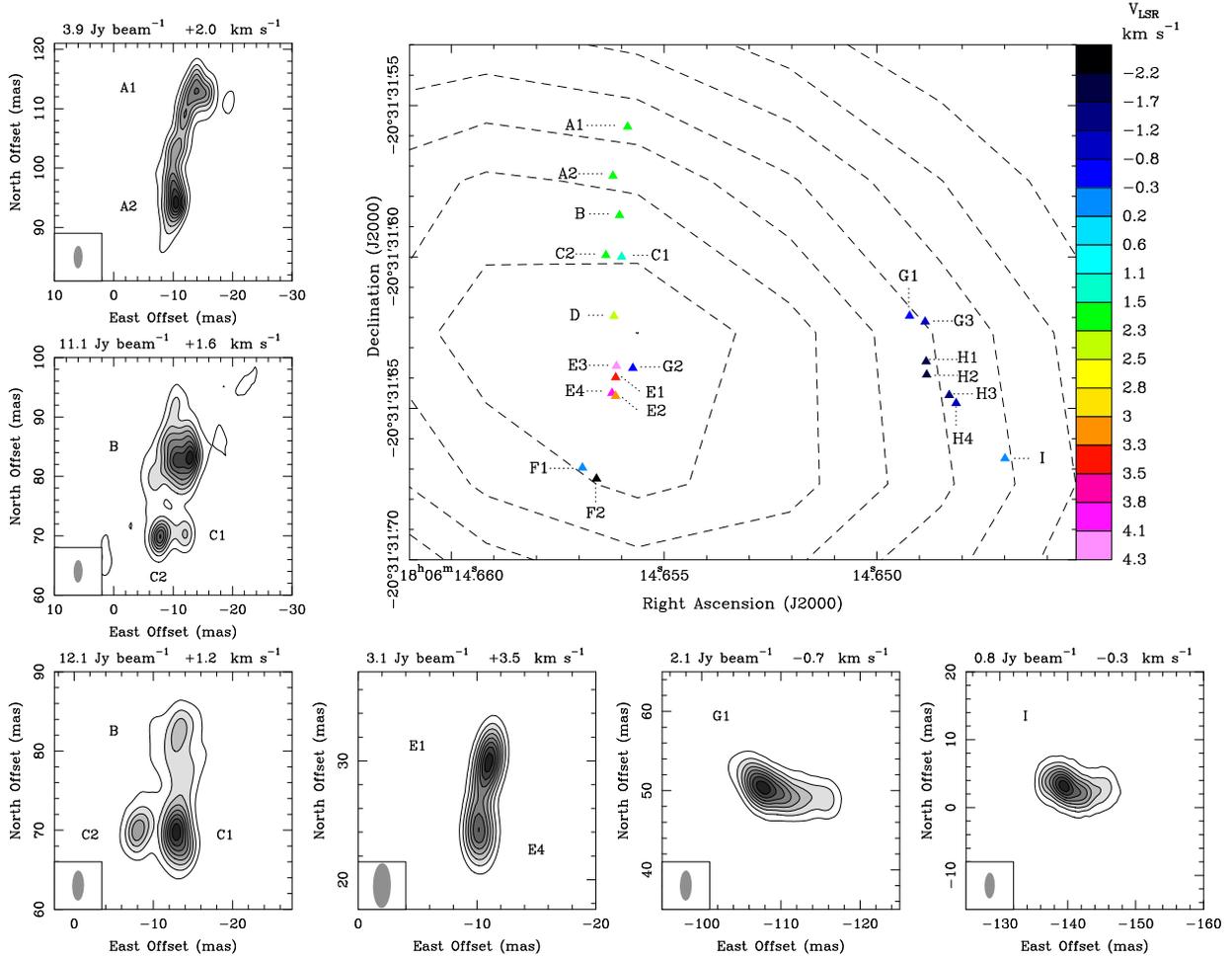}
\caption{G9.62+0.20 12~GHz methanol maser distribution and channel-maps of methanol emission centers. \textit{Large Panel:} distribution of the 12~GHz methanol maser peaks (filled triangles) plotted on the 22~GHz continuum VLA BnA-configuration image (dashed contours) of component G9.62+0.20~E \citep{Testi2000}. The plotted levels of the continuum emission are at multiples of 10\% of the peak brightness of 8.5~mJy~beam$^{-1}$. The synthesized HPBW of the VLA image is $ 0.28 \times 0.18~\rm{arcsec}^2 $ at a P.A. of $ 64\degr $. Different colors are used to indicate the maser LSR velocities, according to the color scale on the right-hand side of the plot. The labels of each maser group follow those introduced by \citet{Minier2002} and \citet{Goedhart2005}. \textit{Edge Panels:} channel-maps of the brightest, extended-structure, maser emission from the fifth epoch (2007 March 16). The plotted levels are at multiples of 10\% of the peak intensity reported on the top of each panel as well the LSR velocity of each channel-map. Positions are relative to the phase reference spot F1. The naturally-weighted synthesized beam, with a HPBW of $ 3.8 \times 1.5~\rm{mas}^2 $ at a P.A. of $ -1\degr $, is shown on the bottom-left corner of each panel.
See the electronic edition of the Journal for a color version of this figure.\label{fig3}}
\end{figure*}

Our measurements allow us to locate the distribution of the individual maser spots with respect to the
HCH~II component \objectname{G9.62+0.20}~E. Figure~\ref{fig3}
presents the absolute positions of the 12~GHz maser spots superposed on the 22~GHz continuum emission map
of \citet{Testi2000}. For this data taken in the BnA-configuration, the VLA has a synthesized HPBW
of $ 0.28 \times 0.18~\rm{arcsec}^2$ at a P.A. of $ 64\degr $ east of north and
is insufficient to resolve the emission of the HCH~II region E.
The accuracy in the absolute position of the VLA map is better than 0.2 arcsec.
The spectral index of component E at radio frequencies, from 8.4 to 110~GHz, is 0.95 \citep{Franco2000}.
We find that individual 12~GHz methanol maser spot velocities and flux densities, as well as their overall
distribution, change little over a time span of about 8~yr, from 1998 November
\citep{Minier2002} to our observations in 2007 March.  We detected all nine distinct clusters imaged by
\citet{Minier2000,Minier2002} within a region of about $ 150 \times 130$~mas$^2$, labeled from A to I,
and resolved them into 20 distinct spots/knots. The maser emission generally appears as compact cores
of high brightness embedded in extended low brightness structures (Figure~\ref{fig3} edge panels).

\begin{figure*}
\centering
\includegraphics[angle= 0, scale= 0.7]{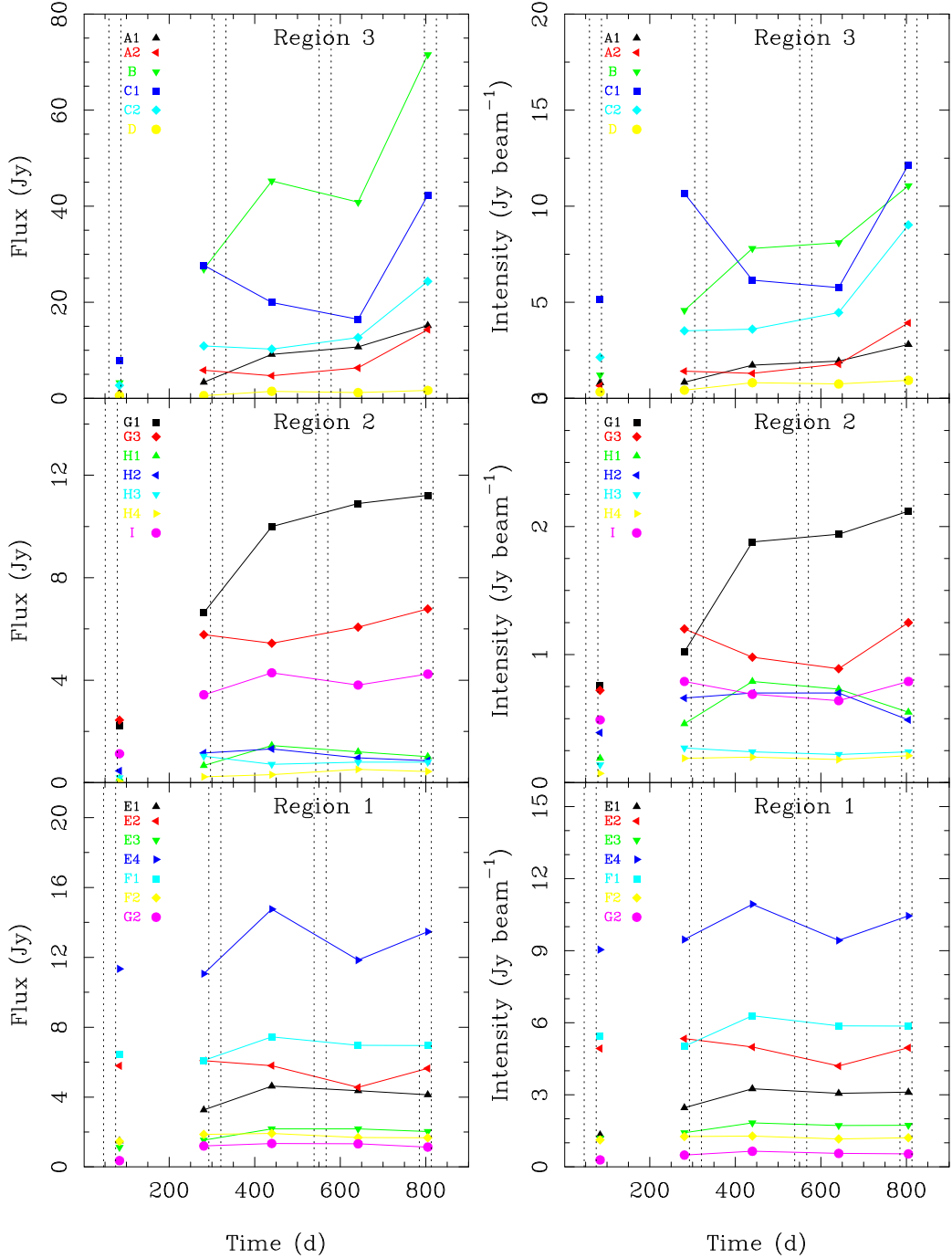}
\caption{Plot of Light Curves of the 12~GHz methanol maser emission in G9.62+0.20~E at our 5 epochs: 2005 March 25 and October 8, 2006 March 16 and October 4, 2007 March 16. Day 0 is 2005 January 1 (JD 2453371). The three panels are divided according to the 3~regions defined by \citet{Goedhart2005}. Dotted lines enclose the central period of each flare, from 3 weeks (start of steep rising) to 7 weeks  after the start of each flare cycle. Timing of flares was evaluated using the ephemeris provided by \citet{Goedhart2003}. For Region 1 and 2 the timing of flares was advanced, with respect to Region 3, of 12 and 8 days, respectively, according to Figure~7 in \citet{Goedhart2005}. Values from the first epoch should be taken as lower limits because of loss of extended emission due to the lack of short-baselines at that epoch. See the electronic edition of the Journal for a color version of this figure.\label{fig4}}
\end{figure*}

Analysis of the light curves of the brightest channel of each spot is consistent with the timing
of the flares measured by \citet{Goedhart2003}. Light curve data of 12~GHz methanol maser emission
in G9.62+0.20~E at our 5 epochs are plotted in Figure~\ref{fig4}. Brightness (Jy beam$^{-1}$)
and integrated flux density (Jy) of the brightest channel of each maser feature are estimated
using the task IMSTAT. Selected BOXes for IMSTAT purposes are the same per spot at every epoch.
Note that flux densities from the first epoch should be taken as lower limits because of the loss of
extended emission due to the lack of short-baselines at that epoch (cf. visibility amplitude vs.
\textit{uv}-distance diagrams in \citealt{Minier2002}).
Epoch~2 falls just before the start of the flare predicted by the periodicity and phasing reported
by \citet{Goedhart2003}; epoch~3 and 4 fall in the expected quiescent periods. Interestingly, Epoch~5
(2007 March) falls during the expected flaring time and our data for features B and C1 (which displayed
the strongest flares previously; cf. Figure~7 in \citealt{Goedhart2005}) show the greatest increases
of flux density.

\clearpage

\clearpage

\clearpage

\begin{deluxetable}{llllrlc}
\tablecaption{Positions and Brightness\label{tab1}}
\tablewidth{0pt}
\tablehead{
\colhead{Source} & \colhead{R.A.~(J2000)} & \colhead{Dec.~(J2000)} & \colhead{$\theta_{sep}$}
& \colhead{P.A.} & \colhead{T$_{b}$} & \colhead{V$_{LSR}$} \\
\colhead{ }       & \colhead{(h m s)}       & \colhead{($\degr$ ' '')}    & \colhead{($\degr$)}
& \colhead{($\degr$)} & \colhead{(Jy beam$^{-1}$)} & \colhead{(\kms)} \\}

\startdata
G9.62+0.20 & 18 06 14.6568 & -20 31 31.670 & \nodata & \nodata  & 5.0--6.3 &  +0.1 \\
J1803--2030 & 18 03 23.7204 & -20 30 17.242 & 0.7     & -70  & 0.011   & \nodata \\
J1811--2055 & 18 11 06.7911 & -20 55 03.286 & 1.2     & 127  & 0.018   & \nodata \\
\enddata
\tablecomments{Absolute positions are accurate to about $\pm$~20~mas. Angular offsets ($\theta_{sep}$) and position angles (P.A.) east of north relative to the maser source are indicated in columns 4 and 5. Brightness (T$_{b}$) for the background sources are from the third epoch. Restoring beam sizes (FWHM) were 1.5~mas (round) for G9.62+0.20 and J1803--2030, and 3.0~mas (round) for J1811--2055.}
\end{deluxetable}

\clearpage

\begin{deluxetable}{lllll}
\tablecaption{G9.62+0.20:~Parallax \& Proper Motion Fit\label{tab2}}
\tablewidth{0pt}
\tablehead{
\colhead{Maser V$_{LSR}$}   & \colhead{Background} & \colhead{Parallax} & \colhead{$\mu_x$} & \colhead{$\mu_y$} \\
\colhead{(\kms)}     & \colhead{Source}     & \colhead{(mas)}    & \colhead{(mas yr$^{-1}$)} & \colhead{(mas yr$^{-1}$)} \\}

\startdata
+0.1 & J1803--2030  & $0.171 \pm 0.026$ & $-0.635 \pm 0.059 $ & $-2.76 \pm 0.11 $  \\
+0.1 & J1811--2055 & $0.218 \pm 0.050$ & $-0.526 \pm 0.091 $ & $-2.22 \pm 0.31 $  \\
    &         &                   &                     &                    \\
 Mean & \nodata & $0.194 \pm 0.023$ &  $-0.580 \pm 0.054 $ & $-2.49 \pm 0.27 $  \\
\enddata
\tablecomments{Column 1 reports the LSR velocity of the reference maser channel; column 2 indicates the background sources whose data were used for the parallax fit; column 3 reports the fitted parallax; columns 4 and 5 give the fitted proper motions along the east and north direction, respectively. ``Mean'' is the un-weighted average of the measurements relative to the two background sources.}
\end{deluxetable}

\end{document}